\documentclass[journal]{IEEEtran}
\usepackage{cite}

\ifCLASSINFOpdf 
\usepackage[pdftex]{graphicx}
\else
\usepackage[dvips]{graphicx}
\fi

\usepackage[cmex10]{amsmath}
\usepackage{algorithmic}
\usepackage[ruled,norelsize]{algorithm2e}

\usepackage{array}

\usepackage{mdwmath}
\usepackage{mdwtab}
\usepackage{todonotes}

\usepackage{eqparbox}

\ifCLASSOPTIONcompsoc
\usepackage[caption=false,font=normalsize,labelfont=sf,textfont=sf]{subfig}
\else
\usepackage[caption=false,font=footnotesize]{subfig}
\fi

\usepackage{fixltx2e}
\usepackage{url}
\usepackage{amsfonts}
\usepackage{amssymb}
\usepackage{amsthm}
\usepackage{bm}
\usepackage[ruled]{algorithm2e}
\usepackage[utf8]{inputenc}
\usepackage{booktabs}
\usepackage{url}
\usepackage{cite}
\usepackage{flushend}
\usepackage{epstopdf}

\begin{document}
%
\title{Toward a Tactile Internet Reference Architecture: Vision and Progress of the IEEE P1918.1 Standard}


\author{
		Adnan~Aijaz,
		Zaher~Dawy,
		Nikolaos~Pappas,
		Meryem~Simsek,
		Sharief~Oteafy,
		and~Oliver~Holland
\thanks{A. Aijaz is with the Telecommunications Research Laboratory, Toshiba Research Europe Ltd., Bristol, BS1 4ND, U.K. 

Z. Dawy is with the Faculty of Engineering and Architecture, American University of Beirut, Lebanon. 

N. Pappas is with the Department of Science and Technology, Link\(\ddot{\text{o}}\)ping University, Sweden. 

M. Simsek is with the International Computer Science Institute Berkeley, Berkeley, CA, USA. 

S. Oteafy is with the School of Computing, DePaul University, Chicago, IL, USA.

O. Holland is with the Department of Informatics, King's College London, London, U.K. 

Contact e-mail: adnan.aijaz@toshiba-trel.com}}
\maketitle
\begin{abstract}
\boldmath
The term \emph{Tactile Internet} broadly refers to a communication network which is capable of delivering control, touch, and sensing/actuation information in real-time. The Tactile Internet is currently a topic of interest for various standardization bodies. The emerging IEEE P1918.1 standards working group is focusing on defining a framework for the Tactile Internet. The main objective of this article is to present a reference architecture for the Tactile Internet based on the latest developments within the IEEE P1918.1 standard. The article provides an in-depth treatment of various architectural aspects including the key entities, the interfaces, the functional capabilities and the protocol stack. A case study has been presented as a manifestation of the architecture. Performance evaluation demonstrates the impact of functional capabilities and the underlying enablers on user-level utility pertaining to a generic Tactile Internet application. 

\end{abstract}



%
\IEEEpeerreviewmaketitle

\section{Introduction}
\IEEEPARstart{W}{ireless} communications is unarguably an indispensable element of modern life. The unprecedented development of wireless technologies has completely transformed the way we perceive the Internet today. The wireless revolution has successfully connected a vast majority of the global population. The focus of wireless is now shifting towards providing ubiquitous connectivity for machines and devices, which will create the Internet-of-Things. Recently, the notion of \emph{Tactile Internet} \cite{Tac_Int} has emerged, which is envisioned to provide a paradigm shift by enabling wireless for real-time steering and control communications. Such powerful wireless connectivity would enable remote interaction for humans, e.g., through exchange of real-time haptic information, or for machines, e.g., through exchange of control and sensing/actuation information in real-time. 

In order to provide a medium for delivering haptic and control information in real-time, the Tactile Internet requires highly reliable, responsive, and intelligent wireless connectivity. The high-availability, the ultra-fast reaction times,
and the carrier-grade reliability of the Tactile Internet will also add a new dimension to human-machine interaction by building real-time interactive systems \cite{ITU_TI}.
The Tactile Internet is envisioned to enable unprecedented applications that will have a marked impact on almost every segment of the society \cite{TI_JSAC}.

In March 2016, the IEEE Standards Association approved the creation of the IEEE P1918.1 standards working group \cite{1918_1}. The scope of the baseline standard is to define a framework for the Tactile Internet, including description of its application scenarios, definitions and terminology, technical assumptions, architecture and reference models, and functional capabilities. The working group is actively engaged in standardizing various aspects of the Tactile Internet since its kick-off meeting in May 2016. The core activities of the working group can be broadly classified into three distinct areas: (i) definitions and use-cases, (ii) haptic codecs, and (iii) reference architectural framework. The  group defines the Tactile Internet as \emph{a network, or a network of networks, for remotely accessing, perceiving, manipulating or
controlling real and virtual objects or processes in perceived real-time}. Some of the key use-cases considered by the working group include teleoperation, immersive virtual reality, interpersonal haptic communication, live haptic broadcasting, automotive, and drone control. The ongoing work on haptic codecs is broadly focused on various aspects of haptic information exchange.

To this end, the main objective of this article is to present a reference architecture for the Tactile Internet in light of the latest developments within the IEEE P1918.1 standard. A preliminary investigation of the architectural aspects of Tactile Internet has been conducted in some  recent studies \cite{aijaz_hc_wcm}, \cite{TI_JSAC}. However, a comprehensive and detailed treatment of the  architecture, including key entities, physical and logical interfaces, and functional description is not available in literature, and therefore, it is the main focus of this article.  


\begin{figure*}
\centering
\includegraphics[scale=0.5]{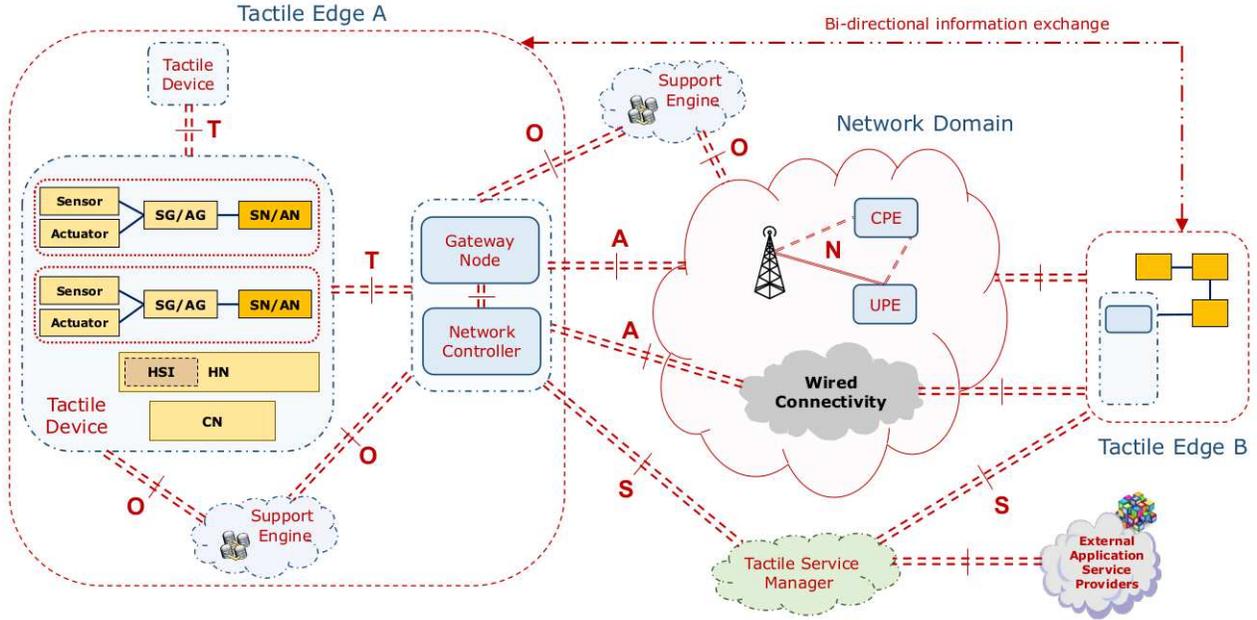}
\caption{The IEEE P1918.1 Tactile Internet reference architecture wherein the gateway node and the network controller are part of the tactile edge.  The key physical interfaces are also shown. }
\label{ref_arch}
\end{figure*}

\section{The IEEE P1918.1 Reference Architecture}
The Tactile Internet has been a topic of interest for various standardization bodies. In August 2014, the ITU released a technology watch report\footnote{The ITU technology watch reports capture new topics for standardization.} \cite{ITU_TI} on the Tactile Internet, identifying it as an important area for innovation. The report also highlighted various use-cases and technical requirements of the Tactile Internet. 
The ETSI IP6 ISG has recently completed a work item on IPv6-based Tactile Internet \cite{ETSI_IP6_TI}. The work item identifies the key features of IPv6 for meeting the stringent technical requirements of the Tactile Internet along with the best practices for different use-cases. 
3GPP has recently released the 5G New Radio (NR) specification \cite{3gpp.38.300} that provides various enhancements pertaining to ultra reliable low latency communication (uRLLC) which is a key enabler for the Tactile Internet.

Development of a reference architecture for the Tactile Internet has  been rarely discussed in any of the aforementioned activities. Therefore, it is one of the key work items in the IEEE 1918.1 standard. To this end, the key principles of the reference architecture design within the IEEE P1918.1 workging group are stated as follows. 

\begin{itemize}
\item To develop a generic architecture that can be mapped to any Tactile Internet application. 

\item To support local area as well as wide area connectivity through wireless or hybrid wireless/wired networking.

\item To have a modular design with flexibility for interworking, computing, caching, intelligence, and other network functions for reliable and responsive service composition. 

\item To minimize dependencies between the device and network domains. 

\item To support separation between control and data planes. 

\item To support integration/interaction with third-party service and application providers. 

\item To enable oversight in resource management to guarantee end-to-end interface mandates.

\item To leverage computing resources from cloud variants at the edge of the  network.
\end{itemize}
 
The IEEE P1918.1 reference architecture for the Tactile Internet, which is illustrated in Fig. \ref{ref_arch}, consists of two tactile edges and a network domain. The tactile edges actually refer to the master and slave domains. The master domain typically consists of a human controller or a machine controller. The slave domain consists of entities which are remotely controlled by the master domain.  The network domain provides the medium for bi-directional information exchange between the tactile edges. Such bi-directional information exchange may result in a closed-loop control system. The network domain can either provide local area connectivity or wide area connectivity through external networks or the Internet. The two tactile edges can also exchange information via peer-to-peer connectivity. Hence,  the tactile edges may also refer to two peer domains.

\begin{figure*}
\centering
\includegraphics[scale=0.5]{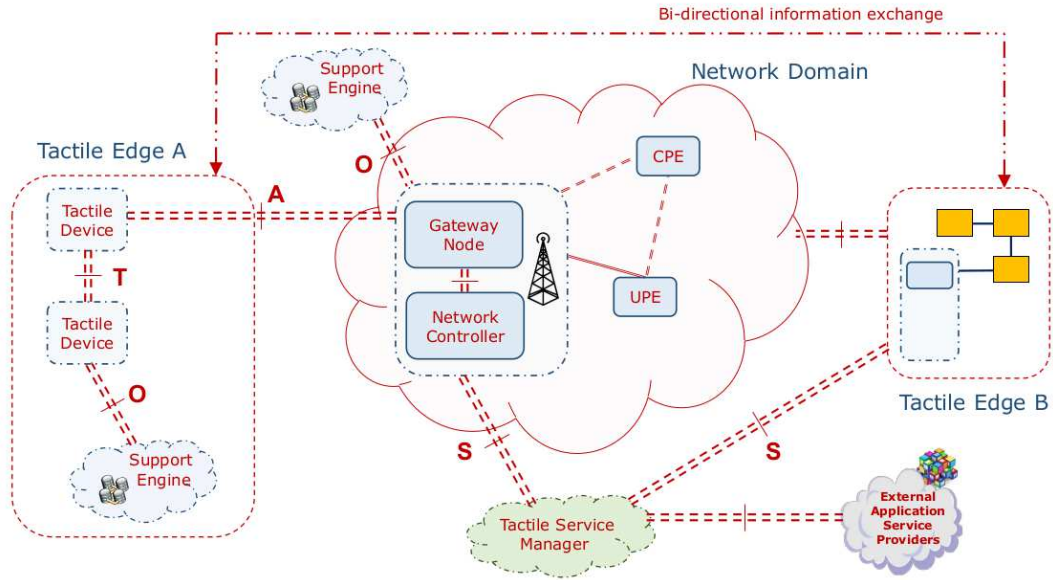}
\caption{An alternate representation of the reference architecture wherein the gateway node and the network controller are part of the network domain.  }
\label{ref_arch2}
\end{figure*}

\subsection{Key Architectural Entities}
The reference architecture consists of a various entities which are described as follows. 
\begin{itemize}
\item \textbf{Tactile Device} -- The tactile device is the core element of any tactile edge. The nature of the tactile device depends on the underlying Tactile Internet application. In one embodiment it consists of a system of sensor nodes (SNs) and actuator nodes (ANs), which are entities with sensing, actuation and limited processing capabilities, respectively, along with necessary connectivity modules for networking capabilities. In a more basic form, third party sensors and actuators can be connected to SNs and ANs, through a sensor gateway (SG) and an actuator gateway (AG), respectively. In another embodiment, the tactile device consists of a human system interface (HSI) that converts human input to haptic input. If equipped with networking capabilities, such a device is referred to as the HSI node (HN). In yet another embodiment, the tactile device consists of a controller which runs control algorithms for controlling a system of SNs and ANs. If equipped with networking capabilities, such a device is referred to as the controller node (CN).

\item \textbf{Gateway Node} -- The gateway node is an entity with enhanced networking capabilities that provides interworking functionality between a tactile edge  and the network domain. The gateway node can be co-located with the network controller (described later) and may exist in the tactile edge or the network domain (see Fig. \ref{ref_arch} and Fig. \ref{ref_arch2}). 

\item \textbf{Network Controller} -- The network controller is an entity that handles the operation of the tactile edge through intelligent network-side and device-side functions. It further comprises of the network management controller (NMC) and device management controller (DMC). The network controller can be co-located with the gateway node, in which case the resulting entity is termed as the gateway network controller. It  may exist in the tactile edge or the network domain.

\item \textbf{Support Engine} -- The support engine refers to an entity that provides computing and/or storage resources for improving the performance/experience of the tactile edge. The support engine can be part of the tactile edges or the network domain or both. In one embodiment, the support engine runs predictive intelligence algorithms for enabling the perception of real-time connectivity under the imperfections of wireless environments. In another embodiment, the support engine provides computation offloading capabilities by handling processing operations that are resource intensive for the tactile devices. In yet another embodiment, the support engine provides caching capability. 

\item \textbf{Tactile Service Manager} -- The tactile service manager is a network domain entity which is primarily responsible for providing interface to external service and application providers. In one embodiment, it may also be responsible for session-level functionalities, e.g., authentication and access rights, session establishment, and charging and billing. 

\item \textbf{User-Plane Entity (UPE)}  -- An entity which handles the user-plane functions for the tactile edge inside the network domain, e.g., context activation, data forwarding to external networks, and quality-of-service (QoS) support.

\item \textbf{Control-Plane Entity (CPE)}  -- An entity which handles the control-plane functions for the tactile edge inside the network domain, e.g., authentication, session establishment, and mobility management.

\end{itemize}

\subsection{Functional Description}
The functional description of the reference architecture is provided as follows.  Note that two variants of the reference architecture can be distinguished depending on the placement of the gateway network controller. In the first scenario (Scenario \(1\)), which is depicted in Fig. \ref{ref_arch}, this entity is part of the tactile edge. The tactile edges A and B represent the master and slave domains, respectively, or two peer domains.  The network domain provides the medium for bi-directional information exchange of control and feedback signals between the tactile devices in the two tactile edges. The network domain may represent wireless connectivity (cellular, Wi-Fi, etc.), wired connectivity (fieldbus system, industrial Ethernet, etc.), or hybrid wired/wireless connectivity. In the second scenario (Scenario \(2\)), which is depicted in Fig. \ref{ref_arch2}, the gateway node and the network controller reside in the network domain. 

The support engine can be realized through different manifestations of the  edge-computing paradigm such as fog-computing, cloudlet-computing, and mobile-edge computing. The support engine can be centralized or distributed in nature. As mentioned earlier, the support engine provides predictive intelligence, computation offloading and/or storage/caching functionalities. For predictive intelligence, the support engine runs a model of the Tactile Internet application which could be obtained in either online or offline manner. The support engine may provide full or partial computation offloading capabilities. The offloading decision is made by either the tactile device or the gateway network controller. 

The tactile service manager is an optional entity. Its existence is dependent on the use-case and the underlying connectivity technology. The user-plane and control-plane entities are specific to the underlying connectivity technology.

Note that the reference architecture mainly encompasses entities which are important from a networking and standardization perspective. There could potentially be other entities that are part of the tactile edges or the network domain. For instance, in some Tactile Internet applications the master tactile edge may have the provisioning for audio/visual feedback from the slave tactile edge. Hence, the tactile edge may additionally contain necessary audio/visual equipment. 

\section{Key Interfaces in the Reference Architecture}
The IEEE P1918.1 working group has adopted the alphabetical naming convention for the interfaces between different architectural entities. The key physical and logical interfaces are described as follows.

\subsection{Key Physical Interfaces}

\begin{itemize}
\item \textbf{Access Interface} -- The access interface, or the A interface, provides connectivity between a tactile edge and the network domain. It is the main reference point for user-plane and control-plane information exchange between the  network domain and the tactile edge. The end points of the A interface are actually dependent on the variant of the reference architecture. In Scenario \(1\), the A interface connects the gateway node and the network controller to a network domain entity such as a cellular base station. In Scenario \(2\), the A interface connects a tactile device to the gateway node and the network controller, which are part of the network domain. 

\item \textbf{Tactile Interface} -- The tactile interface, or the T interface, provides connectivity between entities of the tactile edge. It is the main reference point for user-plane and control-plane information exchange between the entities of the tactile edge. In Scenario 1, it either connects a tactile device to the gateway node and the network controller or provides connectivity between two tactile devices. In Scenario 2, it only provides connectivity between two tactile devices. The T interface can also be based on peer-to-peer connectivity paradigms which are under active investigation within the IEEE 802.15.8-2017 standard \cite{802_15_8}. 

\item \textbf{Open Interface} -- The open interface, or the O interface, provides connectivity between any architectural entity and the support engine.  In on embodiment, it provides connectivity between the support engine and any entity pertaining to the tactile edge, e.g., the tactile device or the gateway node and the network controller. In another embodiment, it provides connectivity between the support engine and any network domain entity such as a base station or an access point. 

\item \textbf{Service Interface} -- The service interface, or the S interface, provides connectivity between the tactile service manager and the gateway node and the network controller. The S interface only carries the control-plane information. 

\item \textbf{Network-side Interface} -- The network-side interface, or the N interface refers to any interface providing internal connectivity between network domain entities. The N interface is the main reference point for user-plane or/and control-plane information exchange between various network domain entities. To further distinguish the N interface, various supplementary interfaces have been defined. The N1 interface provides connectivity between the control-plane entity and the base station (in Scenario \(1\)) or the gateway network controller (in Scenario \(2\)). Similarly, the N2 interface provides the respective connectivity with the user-plane entity. Finally, the N3 interface interconnects the user-plane entity and the control-plane entity. 
\end{itemize}

\subsection{Key Logical Interfaces}
In addition to the above mentioned physical interface, the reference architecture has identified various logical interfaces. 

\begin{itemize}
\item \textbf{L0 Interface} -- The L0 interface interconnects the gateway node and the network controller. 

\item \textbf{L1 Interface} -- The L1 interface interconnects a tactile device and the tactile service manager.

\item \textbf{L2 Interface} -- The L2 interface interconnects a tactile device and the control-plane entity.

\item \textbf{L3 Interface} -- The L3 interface interconnects a tactile device and the user-plane entity. 

\end{itemize}


\section{Functional Capabilities}
To support the envisioned Tactile Internet applications, the interfaces must fulfil certain performance requirements or capabilities. With reference to the majority of Tactile Internet applications, such capabilities can be characterized in terms of various metrics such as availability, latency, reliability, and scalability. In context of Tactile Internet, we define these metrics as follows. 

\begin{itemize}
\item \textbf{Availability} -- The availability of an interface is a measure of its accessibility. It is defined as the probability that a given interface is available for any user-plane or control-plane connectivity service. 

\item \textbf{Reliability} -- The reliability of an interface is a measure of its packet delivery performance. It is defined as the capability of transmitting a fixed-size protocol data unit (PDU) within a predefined time duration with high success probability. 

\item \textbf{Latency} -- The latency of an interface is a measure of its responsiveness. It is defined as the capability to successfully deliver a protocol layer packet from its ingress point (at the transmitter) to the same protocol layer egress point (at the receiver) in order to fulfil the end-to-end latency requirement. The end-to-end latency is defined as the one way delay to successfully deliver an application layer packet from a tactile device in tactile edge A to a tactile device in tactile edge B. 

\item \textbf{Scalability} -- The scalability of an interface describes its capability to cope and perform under an increased number of devices. It is defined as the maximum number of devices that can be supported without compromising the availability, reliability, and latency requirements.

\end{itemize}

The working group has specified two different grades of capabilities for interfaces: an ultra-grade and a normal-grade. The functional capabilities of the A and T interfaces are summarized in \tablename~\ref{func_cap}. 
It can be easily inferred that the desired functional capabilities create a set of stringent requirements that need to be fulfilled irrespective of the underlying connectivity technology. Some of the most important enablers for realizing such functional capabilities are stated as follows. 

\begin{table*}[]
\centering
\caption{Functional Capabilities}
\label{func_cap}
\begin{tabular}{@{}ccc@{}}
\toprule
\textbf{Interface}   & \textbf{Ultra-Grade Capabilities} & \textbf{Normal-Grade Capabilities} \\ \midrule
\textbf{A and T Interfaces} &   \begin{tabular}[c]{@{}l@{}}Availability: ultra-high; \textgreater{}\(99.99999\%\)\\ Reliability: ultra-high; \textgreater{}\(99.999\%\) \\ Latency: ultra-low; \(10\%\) of end-to-end latency (e.g., \(1\) ms) \\ Scalability: medium; \(1\) -- \(50\) tactile devices  \end{tabular}          &     \begin{tabular}[c]{@{}l@{}}Availability: very-high; \textgreater{}\(99.999\%\)\\ Reliability: very-high; \textgreater{}\(99.99\%\) \\ Latency: very-low; \(50\%\) of end-to-end latency (e.g., \(10\) ms) \\ Scalability: high; \(50\) -- \(100\) tactile devices  \end{tabular}           \\
 \bottomrule
\end{tabular}
\end{table*}

\begin{figure*}
\label{comb_prot_stack}
\centering
\subfloat[]{\label{prot1}\includegraphics[scale=0.5]{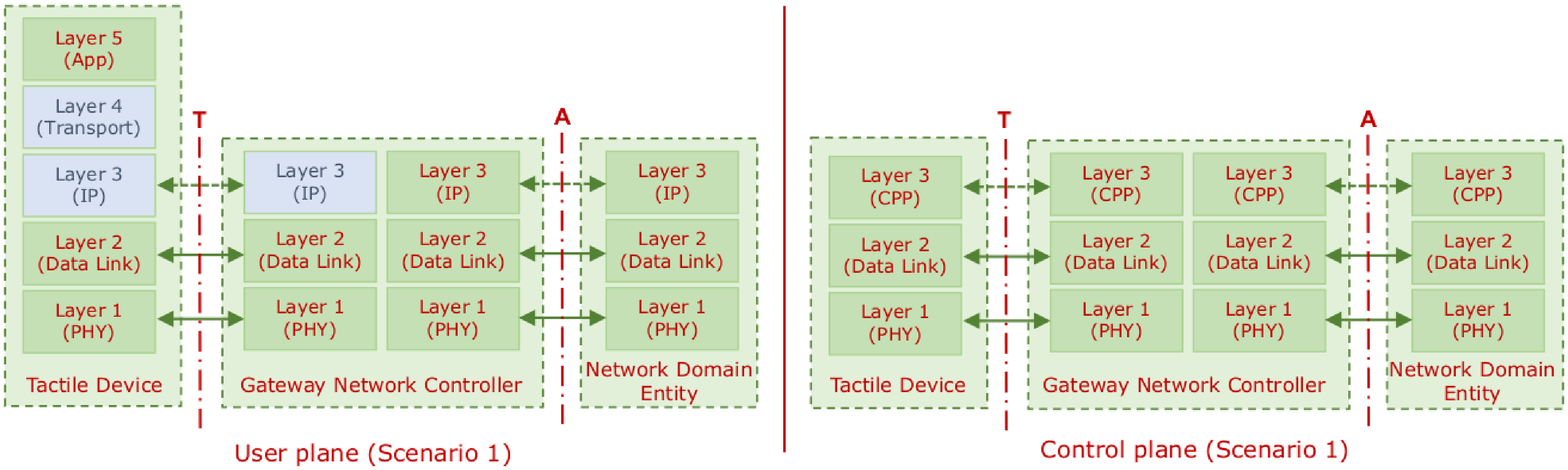}} \
\subfloat[]{\label{prot2}\includegraphics[scale=0.48]{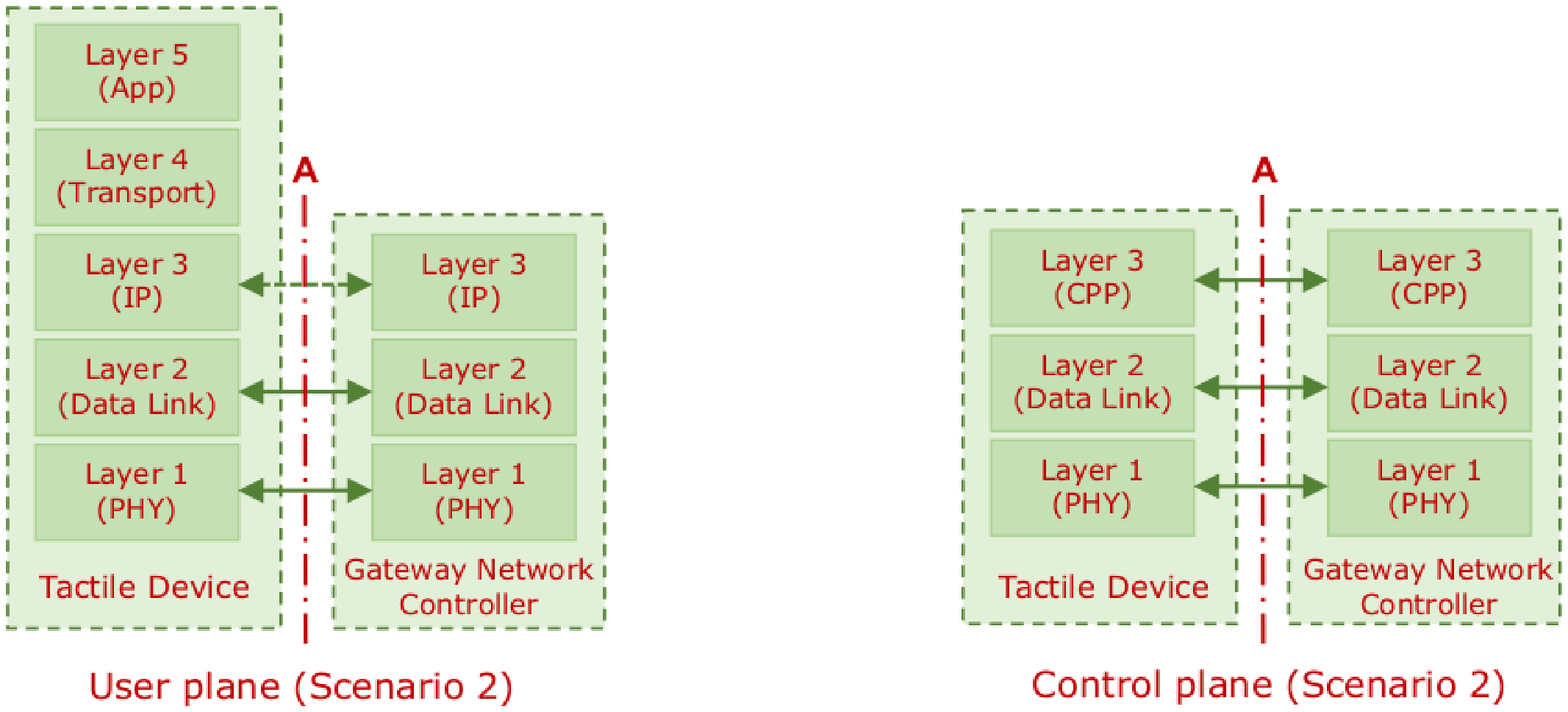}} \
\subfloat[]{\label{prot3}\includegraphics[scale=0.48]{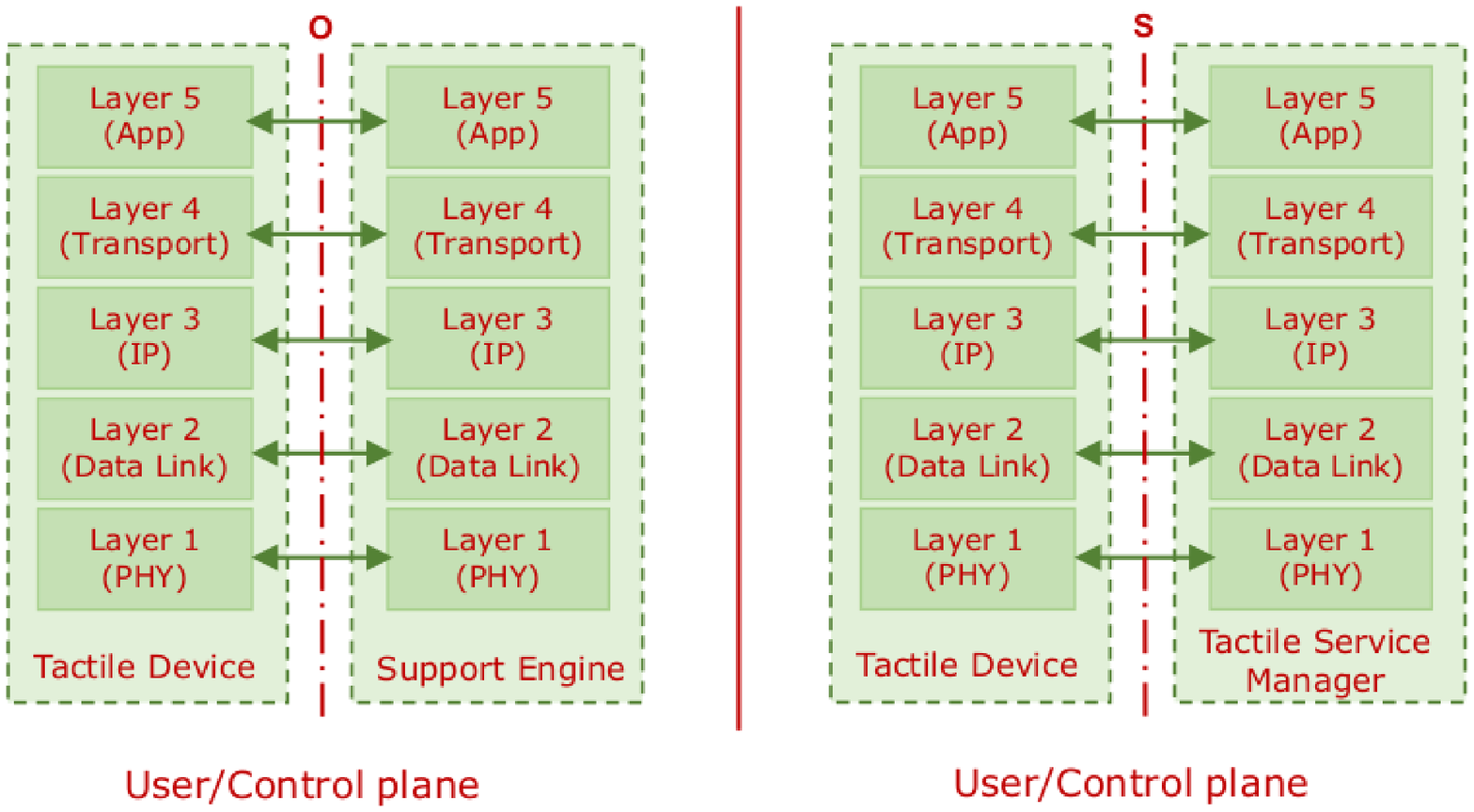}} \
\caption{The user-plane and control-plane protocol stack for different interfaces: (a) the A interface and the T interface for Scenario 1; (b) the A interface in Scenario 2; (c) the O interface and the S interface. The layers highlighted in blue are optional from an implementation perspective.   }
\end{figure*}

\subsection{Key Enablers for High Availability}
Redundancy is the key to achieving high availability. The IEC 62439-3 standard \cite{IEC_PRP}  has specified seamless redundancy protocols for fault tolerance in industrial Ethernet networks. Such protocols can also be applied for achieving high availability in wireless communications.

\subsection{Key Enablers for High Reliability}
Diversity is the key to achieving high reliability. However, diversity in time domain is not suitable as it incurs additional latency. Some of the key enablers for high reliability without incurring additional latency include multi-connectivity, efficient channel coding techniques, cooperative transmissions, and packet duplication with multi-path diversity.

\subsection{Key Enablers for Low Latency}
Achieving low latency becomes particularly challenging, especially considering its trade-off with providing high reliability. Some of the key enablers for achieving low latency, that could be extended to any wireless technology in context of the Tactile Internet, include short packet transmissions, short frame structure, full-duplex communication, lean protocol stack, flexible resource allocation, edge-intelligence, and powerful and efficient hardware designs.

\subsection{Key Enablers for High Scalability}
Achieving high scalability becomes particularly important from network utilization perspective. Some of the key enablers for high scalability include efficient and dynamic multiple access techniques, spectrum aggregation, successive interference cancellation, and interference averaging.

%
%

\section{Overview of Protocol Stack}
The protocol stack for different interfaces is specific to the underlying connectivity technology. In order to provide a generic stack for different interfaces, the working group has adopted the standard TCP/IP protocol stack model. 

Fig. \ref{prot1} depicts the protocol stack for connectivity between the tactile edge and the network domain in Scenario 1, i.e., via the gateway network controller. Note that in this case the gateway network controller has a dual protocol stack. The  stack for the A interface provides connectivity to the network domain whereas the stack for the T interface provides connectivity to the tactile device.  First, we explain the user-plane protocol stack in this case. The application layer (Layer 5) in the tactile device  generates the desired control, haptic, sensing, or actuation information that needs to be transmitted to a tactile device in the other tactile edge. Depending on the nature of the T interface, the tactile device may connect to the gateway network controller at Layer 2 or Layer 3. In the former case, the application layer directly resides over Layer 2. The gateway network controller connects with any network domain entity at Layer 3 over the A interface. From a control-plane perspective, there is Layer 3 connectivity at both T and A interfaces. The control-plane uses the same Layer 1 and Layer 2 as in the user-plane. However, the control-plane protocol (CPP) at Layer 3, which handles typical control-plane functionalities, could be different for each interface. 
 
Fig. \ref{prot2} shows Scenario 2 wherein the tactile device directly connects to the network domain over the A interface. The tactile device connects to the gateway network controller at Layer 3. Note that the latter has a single protocol stack in this case. The user-plane and control-plane protocol stacks for the A interface are similar to that in Scenario 1. 

Fig. \ref{prot3} depicts the protocol stack for the O interface and the S interface. For both interfaces the respective entities hold a full protocol stack for both user-plane and control-plane. The nature of both interfaces is specific to the physical implementation. The protocol stack for various network-side interfaces is specific to the underlying connectivity technology. Hence, it has been deemed as beyond the scope of the working group.

\section{Case Study}
This section  presents a case study that describes how the architecture maps to a specific use-case. We consider the case of multi-player virtual/augmented reality over 5G wireless networks \cite{VR_5G}, which is a prominent application of the Tactile Internet. In this case, the two Tactile edges are connected via a 5G wireless network and correspond to an indoor gaming arcade with multiple players and a virtual/augmented reality server, respectively. The TDs possibly refer to wireless head-mounted virtual/augmented reality displays. The TDs can directly connect to the 5G network through the A interface which would be based on 5G air-interface specifications \cite{3gpp.38.300}. The gateway network controller corresponds to the gNB as part of the 5G radio access network. It provides various functionalities including connection management, radio resource management and mobility management as per 5G specifications. The support engine provides computation offloading functionality for various processing-hungry tasks such as rendering and processing of high-definition video frames. It can also provide local content caching for enhancing the immersive experience and avoiding cybersickness. The tactile service manager provides an interface to various third-party virtual/augmented reality content providers.

\begin{table}
\caption{Parameters for Performance Evaluation}
\label{table1}
\vspace{-3ex}
\begin{center}
\begin{tabular}{ll}
\toprule
\bf{Parameter} & \bf{Value} \\
\midrule
Overall resource pool  & \(100\) Resource Blocks\\
Path loss  & $128.1+37.6\text{log}_{10}(r_{\text{km}})$\\
Standard deviation of Shadowing  & $8$ dB \\
Noise figure & $5$ dB\\
Transmit power (macrocell)  & $36$ dBm \\
Transmit power (small cell)  & $25$ dBm \\
Transmit power (user)  & $18$ dBm \\
Latency requirement & $5$ ms \\
\hline
\end{tabular}
\end{center}
\end{table}

\section{Performance Evaluation}
\begin{figure}
\label{comb_perf}
\centering
\subfloat[]{\label{util_funcs}\includegraphics[scale=0.18]{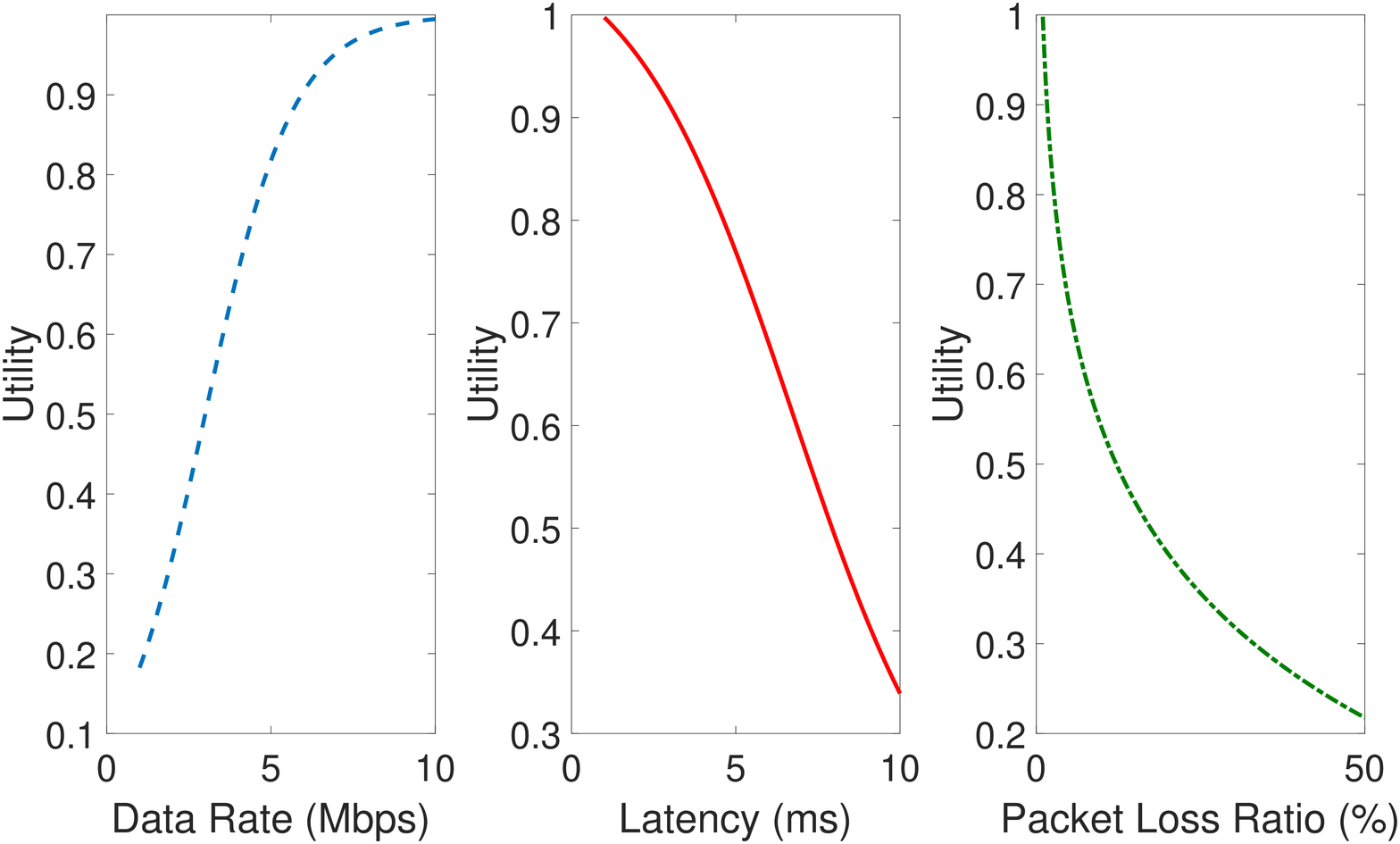}} \
\subfloat[]{\label{sum_util}\includegraphics[scale=0.18]{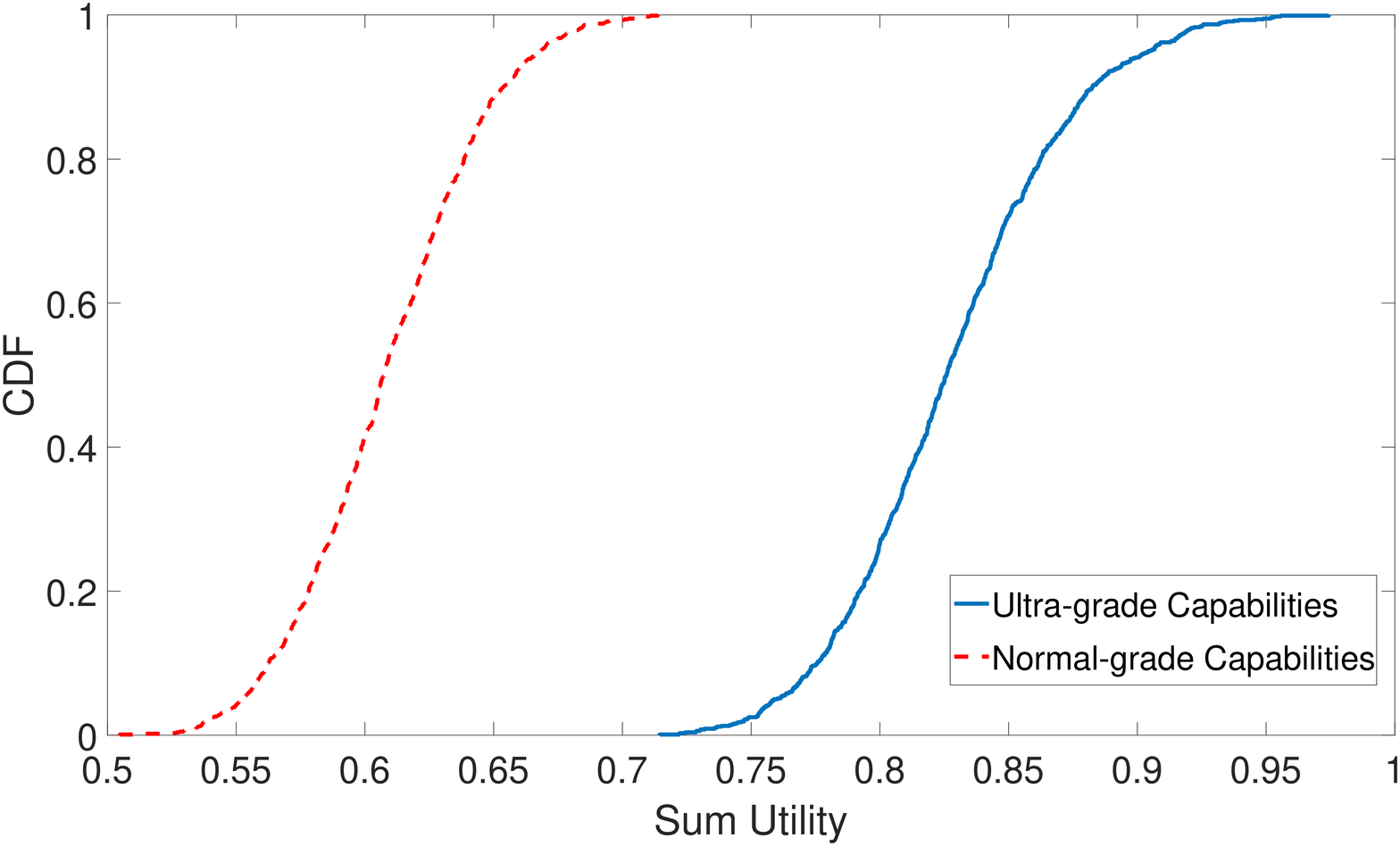}} \
\caption{Performance evaluation for a generic Tactile Internet application: (a) utility functions for data rate, latency and packet loss ratio; (b) CDF of sum utility generated over \(100\) iterations.   }
\end{figure}

To demonstrate the impact of functional capabilities as part of the reference architecture, we conduct a system-level simulation for a generic Tactile Internet application running over a 5G wireless network. We consider a two-tier deployment wherein the first tier consists of a single macrocell with a radius of \(100\) meters. The second tier contains \(4\) small cells, with a radius of \(30\) meters each, randomly distributed in the coverage of the macrocell. We assume \(50\) uniformly distributed users (tactile devices) in the coverage of the macrocell. The channel model accounts for large-scale path loss and shadowing and small-scale Rayleigh fading. Owing to bi-directional information exchange, we assume that the overall utility of a user is jointly determined by its downlink and uplink utilities. Further, the utility in either downlink or uplink is determined by the achieved data rate, the overall latency, and the packet loss ratio. The respective utility functions, which are shown in Fig. \ref{util_funcs}, have been adopted from \cite{utility_mm}.  For the normal-grade capabilities, we assume single-connectivity with macrocell only. For the ultra-grade capabilities, we assume dual-connectivity  solution wherein the user is simultaneously served from two base stations. A user is configured with dual-connectivity only if it is within the coverage of both tiers. Further, we implement packet duplication \cite{aijaz_pd_wcm} using the split bearer architecture in dual-connectivity. The link-level model is based on standard signal-to-noise ratio (SNR). We perform Monte Carlo simulations over different user and small cell distributions with \(1000\) packets (per user) in each iteration. We adopt the joint radio resource allocation for sum utility maximization from \cite{aijaz_twc}. Other simulation parameters are given in \tablename~\ref{table1}.

The results in Fig. \ref{sum_util} show the cumulative distribution function (CDF) of sum utility for all the Tactile Internet users. As shown by the results, ultra-grade capabilities outperform the normal-grade capabilities in terms of meeting the requirements of Tactile Internet application. The performance gain of ultra-grade capabilities is mainly from  packet duplication in dual-connectivity which not only enhances the data rate but also reduces packet losses.

\section{Security Considerations}
Security is an important requirement for many Tactile Internet applications. 
In order to achieve both security and stringent latency requirements, novel approaches are needed. Typically, IP security functionalities are far from the tactile edges, and therefore, providing security with end-to-end latency constraints becomes particularly challenging.   Potential locations/interfaces that must be equipped with security capabilities need to be identified and designed appropriately. Solutions from software-defined networking (SDN) and network function virtualization (NFV) could be adopted, since a thin core network can potentially decrease protocol overheads and reduce latency.

 The security considerations are often specific to the scenario or application. However, partially these need to be addressed at the architectural and protocol level. In the proposed architecture, the interface that provides connection to the external networks/interfaces can contain firewall protection features. In addition, if the communication of internal elements is through wireless links then eavesdropping must be considered as a threat and measures need to be taken. For instance, malicious users/hackers may want to invade the system in battlefields or in cases where remote surgeries are taking place.  It is  also important to identify the modules that must support security, such as end-to-end encryption of the critical data between paired end devices, especially against man-in-the-middle attacks. Security at the PHY layer is also attractive as it does not require and additional layer incurring overhead and latency. 
 
 

The current architecture supports a registration phase for critical components in the architecture, that will aid in rapid authentication and careful augmentation of resources/devices.

\section{Concluding Remarks}
The Tactile Internet has received significant attention from various standardization bodies. This article presented a reference architecture for the Tactile Internet based on the latest developments within the IEEE P1918.1 standards group. The reference architecture has been characterized in terms of various entities and interfaces. Further, functional capabilities for different interfaces have been identified which are crucial in providing the required QoS for the Tactile Internet application as demonstrated by the performance evaluation. Such capabilities would be realized through various technology enablers at different layers of the protocol stack.  The reference architecture has been developed in a connectivity-agnostic manner. Hence, it can be mapped to any Tactile Internet application over any connectivity technology include 5G.




\bibliographystyle{IEEEtran}

\bibliography{IEEEabrv,arch_bib}

\begin{thebibliography}{10}
\providecommand{\url}[1]{#1}
\csname url@samestyle\endcsname
\providecommand{\newblock}{\relax}
\providecommand{\bibinfo}[2]{#2}
\providecommand{\BIBentrySTDinterwordspacing}{\spaceskip=0pt\relax}
\providecommand{\BIBentryALTinterwordstretchfactor}{4}
\providecommand{\BIBentryALTinterwordspacing}{\spaceskip=\fontdimen2\font plus
\BIBentryALTinterwordstretchfactor\fontdimen3\font minus
  \fontdimen4\font\relax}
\providecommand{\BIBforeignlanguage}[2]{{%
\expandafter\ifx\csname l@#1\endcsname\relax
\typeout{** WARNING: IEEEtran.bst: No hyphenation pattern has been}%
\typeout{** loaded for the language `#1'. Using the pattern for}%
\typeout{** the default language instead.}%
\else
\language=\csname l@#1\endcsname
\fi
#2}}
\providecommand{\BIBdecl}{\relax}
\BIBdecl

\bibitem{Tac_Int}
G.~Fettweis, ``{The Tactile Internet: Applications and Challenges},''
  \emph{IEEE Veh. Technol. Mag.}, vol.~9, no.~1, pp. 64--70, 2014.

\bibitem{ITU_TI}
\BIBentryALTinterwordspacing
ITU-T, ``{The Tactile Internet},'' {International Telecommunication Union
  (ITU)}, Technology Watch Report, August 2014. [Online]. Available:
  \url{https://www.itu.int/dms_pub/itu-t/oth/23/01/T23010000230001PDFE.pdf}
\BIBentrySTDinterwordspacing

\bibitem{TI_JSAC}
M.~Simsek, A.~Aijaz, M.~Dohler, J.~Sachs, and G.~Fettweis, ``{5G-Enabled
  Tactile Internet},'' \emph{IEEE J. Sel. Areas Commun.}, vol.~34, no.~3, pp.
  460--473, March 2016.

\bibitem{1918_1}
\BIBentryALTinterwordspacing
``{1918.1 - Tactile Internet: Application Scenarios, Definitions and
  Terminology, Architecture, Functions, and Technical Assumptions}.'' [Online].
  Available: \url{https://standards.ieee.org/develop/project/1918.1.html}
\BIBentrySTDinterwordspacing

\bibitem{aijaz_hc_wcm}
\BIBentryALTinterwordspacing
A.~Aijaz, M.~Dohler, A.~H. Aghvami, V.~Friderikos, and M.~Frodigh, ``{Realizing
  the Tactile Internet: Haptic Communications over Next Generation 5G Cellular
  Networks},'' \emph{IEEE Wireless Commun.}, 2015. [Online]. Available:
  \url{http://arxiv.org/abs/1510.02826}
\BIBentrySTDinterwordspacing

\bibitem{ETSI_IP6_TI}
ETSI, ``{IPv6-based Tactile Internet},'' {International Telecommunication Union
  (ITU)}, Group Specification GR IP6 0014, 2017.

\bibitem{3gpp.38.300}
\BIBentryALTinterwordspacing
3GPP, ``{NR and NG-RAN Overall Description},'' {3rd Generation Partnership
  Project (3GPP)}, TS {38.300}, Dec. 2017, {v2.0}. [Online]. Available:
  \url{http://www.3gpp.org/ftp/Specs/archive/38_series/38.300/}
\BIBentrySTDinterwordspacing

\bibitem{802_15_8}
\BIBentryALTinterwordspacing
``{IEEE Standard for Wireless Medium Access Control (MAC) and Physical Layer
  (PHY) Specifications for Peer Aware Communications (PAC)},'' \emph{IEEE Std
  802.15.8-2017}, 2017. [Online]. Available:
  \url{https://standards.ieee.org/findstds/standard/802.15.8-2017.html}
\BIBentrySTDinterwordspacing

\bibitem{IEC_PRP}
\BIBentryALTinterwordspacing
IEC, ``{Industrial Communications Networks -- High Availability Automation
  Networks -- Part 3: Parallel Redundancy Protocol (PRP) and High-Availability
  Seamless Redundancy (HSR)},'' {International Electrotechnical Commission
  (IEC)}, Standard (2nd Edition) {62439-3}, 2012. [Online]. Available:
  \url{https://webstore.iec.ch/publication/24447}
\BIBentrySTDinterwordspacing

\bibitem{VR_5G}
\BIBentryALTinterwordspacing
M.~S. ElBamby, C.~Perfecto, M.~Bennis, and K.~Doppler, ``{Towards Low-Latency
  and Ultra-Reliable Virtual Reality},'' 2018. [Online]. Available:
  \url{http://arxiv.org/abs/1801.07587}
\BIBentrySTDinterwordspacing

\bibitem{utility_mm}
M.~Mu, A.~Mauthe, and F.~Garcia, ``{A Utility-Based QoS Model for Emerging
  Multimedia Applications},'' in \emph{Next Generation Mobile Applications,
  Services and Technologies (NGMAST)}, Sept 2008, pp. 521--528.

\bibitem{aijaz_pd_wcm}
\BIBentryALTinterwordspacing
A.~Aijaz, ``{Packet Duplication in Dual Connectivity Enabled 5G Wireless
  Networks: Overview and Challenges},'' \emph{IEEE Wireless Commun.}, 2018.
  [Online]. Available: \url{https://arxiv.org/abs/1804.01058}
\BIBentrySTDinterwordspacing

\bibitem{aijaz_twc}
------, ``{Toward Human-in-the-Loop Mobile Networks: A Radio Resource
  Allocation Perspective on Haptic Communications},'' \emph{IEEE Trans.
  Wireless Commun.}, vol.~17, no.~7, pp. 4493--4508, 2018.

\end{thebibliography}
%

\end{document}